\documentclass[
    prx,
    reprint,
    amsmath,amssymb,
    a4paper,
    10pt,
    superscriptaddress,
    nofootinbib
]{revtex4-2}

\usepackage{graphicx}
\graphicspath{{fig/}}
\usepackage[dvipsnames,table,xcdraw]{xcolor}
\definecolor{rmpblue}{HTML}{2e3092}
\usepackage[
	colorlinks=true,
	citecolor=rmpblue,
	linkcolor=rmpblue,
	filecolor=magenta,
	urlcolor=rmpblue,
]{hyperref}
\usepackage{physics}
\usepackage{siunitx}

\newcommand{\qqoute}[1]{`#1'}

\newcommand{\ort}[1]{\bar{\boldsymbol{\mathbf{#1}}}}

\newcommand{\iu}{\mathrm{i}\mkern1mu}

\newcommand{\affilANU}{Research School of Physics, Australian National University, Canberra ACT 2601, Australia}
\newcommand{\affilITMO}{School of Physics and Engineering, ITMO University, St.~Petersburg 191002, Russia}
\newcommand{\affilETH}{Nanophotonic Systems Laboratory, Department of Mechanical and Process Engineering,\\ ETH Z\"urich, 8092 Z\"urich, Switzerland}

\begin{document}
	
\title{Optical Supertorque Induced by Mie-Resonant Modes}

\author{Ivan Toftul}
\email{ivan.toftul@anu.edu.au}
\affiliation{\affilANU}

\author{Mihail Petrov}
\affiliation{\affilITMO}

\author{Romain Quidant}
\affiliation{\affilETH}

\author{Yuri Kivshar}
\email{yuri.kivshar@anu.edu.au}
\affiliation{\affilANU}

\begin{abstract}
We introduce the concept of \textit{resonant optical torque} that allows enhancing substantially a transfer of optical angular momentum (AM) of light to a subwavelength particle. We consider high-index cylindrical dielectric nanoparticles supporting Mie resonances, and explore a transfer of AM and how it is affected by absorption and particle shape. We analyze a simple trapping geometry of standing wave patterns created by opposite helical light waves. We uncover \textit{stable} rotation of particles in both nodes and anti-nodes, 
and also study how specific particle properties influence the resonant optical torque. 
We demonstrate that adjusting particle asymmetry and losses can maximize spinning torque, and we predict ``\textit{supertorque}'' originating from the scattering channel mixing.
Our study offers a deeper understanding of the physics of resonant optical torque and its importance in manipulating AM transfer in optical systems, with promising implications for various fields and inspiring further research in resonant light-matter interactions.
\end{abstract}

\maketitle

\section{Introduction} 

Manipulation of particles using light has been a subject of extensive research and exploration in the realm of optics and photonics~\cite{Shi2022Sep,Zhang2008Apr,Fazal2011Jun,Pesce2020Dec}.
Besides manipulation of particles' center of mass, great  interest lies in the area of rotational degrees of freedom as well~\cite{Toftul2023PRL,Wu2022May,Shishkin2020Jan,Andren2021Sep,vanderLaan2020arXiv,Zielinska2023PRL}.
The phenomenon of optical rotation, which has garnered significant attention in recent years, directly relates  the mechanism of angular momentum transfer from light to objects~\cite{Rahimzadegan2017PRB,Liu2010NN,Achouri2023ACS}. There are  several main applications of optical rotation of micro- and nanoobjects in fluid actuation~\cite{Mohanty2012LabChip}, microrobotics~\cite{Hou2023LabChip,Ali2024arXiv}, and for studying fundamental physical concepts~\cite{Reimann2018PRL}.

In optofluidics, where light-induced actuators and motors find their applications in solution,  a significant progress has been achieved in the recent years. 
In particular, rotation of plasmonic nanorods has been extensively studied and used to actuate nanospecimens or their surrounding fluid~\cite{Shao2018Jun, Karpinski2020Sep, Liu2010Aug}, or also for realizing   nanomixer systems \cite{canos_valero_nanovortexdriven_2020}. 
Optical rotation of small non-resonant nanoparticles in vacuum allows to achieve extreme values of optical rotation frequencies.  In the pioneering work~\cite{Arita2013Aug} authors introduced the concept of a microgyroscope, showcasing stable rotation rates of up to $5~\text{MHz}$ and the consequential positional stabilization, resulting in effective cooling of the particle to $40~\text{K}$. Later, in Refs.~\cite{Reimann2018PRL,Ahn2018PRL,Zielinska2024PRL} it was shown that GHz rate of rotation of nanoparticles was achieved by reducing the pressure below $10^{-5}~\text{mbar}$. The combination of high rotation frequencies in the GHz regime with ultrasensitive optical readout makes rotational modes of levitated nanoparticles in vacuum highly relevant to ultrasensitive torque sensing~\cite{Ahn2020Feb} and probing quantum frictions~\cite{Manjavacas2010Sep}.

\begin{figure}
    \centering
    \includegraphics[width=\linewidth]{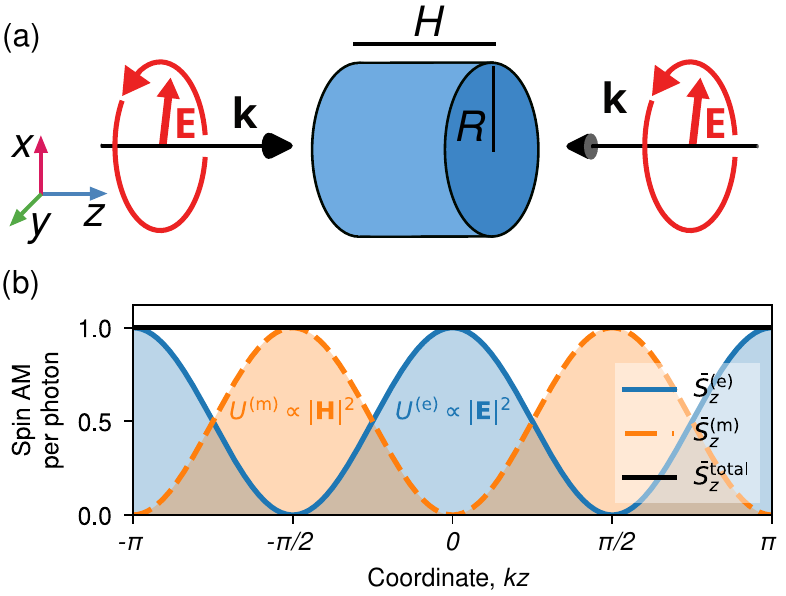}
    \caption{\textbf{ Resonant optical rotation.} (a) Proposed trapping geometry: a high index cylinder is trapped in a standing wave constructed by two counter-propagating beams with opposite helicities. \textbf{(b)} Properties of the standing wave. On the upper half there is a spin angular momentum density distribution (electric $\vb{S}^{(\text{e})}_z$ and magnetic $\vb{S}_z^{(\text{m})}$) normalized to represent values per one photon based on Eq.~\eqref{eq:SAM_standing_wave}. The shaded area shows the electric and magnetic energy densities, $U^{(\text{e})}$ and $U^{(\text{m})}$. }
    \label{fig1:2beams_am}
\end{figure}

Optical torque acting on a scatterer in the electromagnetic field is fundamentally tied to the angular momentum (AM) conservation law and arises due to the imbalance between the incident and rescattered  AM of the field~\cite{Toftul2024arXiv,Sadowsky1899,Poynting1909, Beth1935, Beth1936, Holbourn1936, Brasselet2023AP,Friese1998Nat}. 
When an incident field carries non-zero AM, two primary mechanisms drive this imbalance:
(i) the absence of the rotational symmetry in the particle’s geometry and/or optical properties~\cite{Brasselet2009Jun,Friese1998Jul,Simpson2007Oct,Trojek2012Jul} and (ii) the presence of absorption~\cite{Marston1984Nov,Nieto-Vesperinas2015OL, Nieto-Vesperinas2015PRA}.
In the vast majority of the existing works and discussed applications the nanoparticles and nanostructures were in the non-resonant regimes, leaving the  mechanisms of the resonant increase of optical torque out of consideration and clearly understudied. Recently, the enhancement of optical forces with optical resonances of subwavelength objects started to attract the interest of the researchers \cite{Kiselev2020Sep, Lepeshov2023Jun} also in the prospective  of optical torques and optical rotation \cite{Hakobyan2014Aug, Chen2014SR, Shi2022Feb}. While many years have passed since the first mentions of resonant optical trapping effects   in atoms and dielectric structures  \cite{Ashkin1978Mar, Ashkin1977Jun, Chu1985Jul}, today one of the topical problem is achieving of optomechanical control of resonant nanostructures such as dielectric Mie nanoparticles \cite{Achouri2023ACS, Lepeshov2023Jun, Toftul2023PRL}. 

Within this paper, we uncover the mechanisms of optical rotation and angular momentum transfer from optical beam to  dielectric nanostructures possessing pronounced Mie resonances.  We focus on  maximizing  the optical torque  acting on   \textit{anisotropic} and  \textit{absorptive} nanoparticles addressing two main mechanisms of  angular momentum transfer correspondingly. The Mie resonances have already been actively employed for increasing optical absorption \cite{Miroshnichenko2018PRL} and even reaching regimes of super-scattering   \cite{CanosValero2023NC, Krasikov2021PhysRevAppl} in simple cylindrical structures. 
However, strong resonant scattering  prevents trapping of Mie particles in the common optical tweezer geometry since the subtle trapping condition is ruined by the increased scattering pressure force \cite{Kislov2021APR,Toftul2024arXiv}. Currently there are no rigorous approaches to stable trapping  of Mie particle in a single beam geometry while two  counter propagating coherent beams forming a standing wave  cancel the scattering force and make stable trapping possible \cite{Lepeshov2023Jun, juan_cooperatively_2017, mao_switchable_2025}.
Thus, in this work we   consider trapping of  high-index nanoscale cylinders  in the nodes or anti-nodes of a standing wave. The standing wave is formed by the two counter-propagating beams with opposite helicities as shown in Fig.~\ref{fig1:2beams_am}. This configuration allows stable trapping while canceling the optical pressure force which is irrelevant to the physics of rotation of achiral particles. We predict that for Mie resonant structures of complex shape one can expect breaking the limit of the maximal optical torque expected for spherical particles \cite{Rahimzadegan2017PRB}, thus, reaching ``\textit{supertorque}'' regime.
The effect of the supertorque aligns with the broader class of ``super'' effects, such as super-scattering~\cite{Ruan2011APL,CanosValero2023NC} and super-absorption~\cite{Ladutenko2015N}, where carefully engineered resonances or symmetry breaking lead to performance surpassing conventional expectations, usually for sphere. 
Here, we adopt this naming convention.
The proposed approach may push the limits of the efficiency of current nanorotor systems reaching ultrahigh rotation rates far beyond GHz under reasonable laser intensities.

\section{Results and Discussion}

\subsection{Multipolar expansion}

The problem of  light scattering  on objects comparable or smaller than a wavelength can be efficiently solved using the \textit{multipolar approach}. 
The latter approach has already shown to be well suited for analyzing linear and nonlinear scattering in nanophotonics and designing the optical scatterers with predefined properties~\cite{Babicheva2024AOP,Koshelev2021ACSPhot,Toftul2023PRL,Igoshin2024arXiv,Poleva2023PRB,Rahimzadegan2022AOM,Krasikov2021PRAppl,Kislov2021APR}. Below we apply it to express the relation between multipolar content of the incident and scattered fields, and generated optical torque.

Throughout our study, we focus on monochromatic fields with the frequency $\omega$. The time-averaged optical torque equals the rate of change of the total angular momentum, which can  be written as the integration of the angular momentum flux tensor $\tensor{\mathcal{M}}$ through the closed surface around the scatterer as~~\cite{Toftul2024arXiv,jackson1998ClassicalElectrodynamics,novotny2010PrinciplesNanoOptics,bliokh2014ExtraordinaryMomentumSpin}
\begin{equation}
	\vb{T} = - \oint\limits_\Sigma \tensor{\mathcal{M}} \cdot \vb{n} \dd \Sigma ,
	\label{eq:torque_av}
\end{equation}
where $\vb{n}$ is the outer unit normal vector to the $\Sigma$ surface, $\tensor{\mathcal{M}} = \vb{r} \times \tensor{\mathcal{T}}$, where the linear momentum flux tensor is 
$\tensor{\mathcal{T}} = -  \frac{1}{2} \Re \left[ \varepsilon  \vb{E}^* \otimes \vb{E} + \mu  \vb{H}^* \otimes \vb{H} - \frac{\hat{\vb{I}}}{2}  \left( \varepsilon  \abs{\vb{E}}^2 + \mu  \abs{\vb{H}}^2\right) \right]$  with $\vb{E}$ ($\vb{H}$) being the electric (magnetic) fields, and $\varepsilon$ ($\mu$) are the absolute dielectric (magnetic) permittivity of the surrounding media. 
Total fields can be written as a sum of incident and scattered fields as
\begin{equation}
    \vb{E} = \vb{E}_{\text{inc}} + \vb{E}_{\text{sc}}, \qquad 
    \vb{H} = \vb{H}_{\text{inc}} + \vb{H}_{\text{sc}}.
    \label{eq:inc_plus_sc}
\end{equation}
We apply a  multipolar expansion which generally reads as
\begin{align}
    \sqrt{\varepsilon}\vb{E} &= \mathcal{A} \sum \limits_{mj} (A_{mj} \vb{N}_{mj} + B_{mj} \vb{M}_{mj}), \nonumber \\
    \sqrt{\mu}\vb{H} &= - \iu  \mathcal{A} \sum_{mj} (B_{mj} \vb{N}_{mj} + A_{mj} \vb{M}_{mj} ),
    \label{eq:decomsition_general}
\end{align}
where $\vb{N}_{mj}$ ($\vb{M}_{mj}$) are the electric (magnetic) vector spherical harmonics (or multipoles), $\sum_{mj} \equiv \sum_{j=1}^{\infty} \sum_{m = -j}^{j}$, and $\mathcal{A}$ is the dimensional amplitude coefficient, $[\mathcal{A}] = [\sqrt{\varepsilon}\vb{E}] = [\sqrt{\mu}\vb{H}]$. In Eq.~\eqref{eq:decomsition_general} we have used the relation between vector spherical harmonics (VSH) (see Appendix~\ref{app:VSH}) and the Maxwell curl equation $\vb{H} = (\iu \omega \mu)^{-1} \curl \vb{E}$. Depending on the boundary conditions, one should choose different radial dependencies, i.e. function $z_j (kr)$ in Appendix~\ref{app:VSH}.
The VSH basis is particularly convenient for our analysis, as it provides a well-defined total angular momentum for each multipole.  Each harmonic has a total angular momentum $j$  and an angular momentum projection $m$ along the $z$-axis, both measured in units of $\hbar$.
This becomes apparent once the relationship between VSH and spherical tensors is established (see Supplementary Material in Ref.~\cite{Toftul2023PRL}).
We choose the integration surface $\Sigma$ to be a sphere of radius $r$. Coefficients can be calculated by knowing only radial components of the scattered fields as~\footnote{Connection between $A$ and $B$ and coefficients in multipole decomposition used by J.D. Jackson~\cite{jackson1998ClassicalElectrodynamics} is $A_{mj} = - a_{E}(l=j, m=m) / \sqrt{j(j+1)}$ and $B_{mj} = \iu a_{M}(l=j, m=m) / \sqrt{j(j+1)}$.}
\begin{align}
\begin{split}
    A_{mj} &= \frac{kr}{j(j+1) z_j(kr)} \int \limits_{4\pi} \dd \Omega Y_{mj}^{*} (\mathbf{e} \cdot \vu{r}) , \\
    B_{mj} &= \frac{\iu kr}{j(j+1) z_j(kr)} \int \limits_{4\pi} \dd \Omega Y_{mj}^{*} (\mathbf{h} \cdot \vu{r}) ,
\end{split}
    \label{eq:AmjBmj}
\end{align}
where $\vu{r} = \vb{r}/r$  is the outer unit vector, $k = \sqrt{\varepsilon \mu} \omega$, $Y_{mj}(\vartheta, \varphi)$ is the scalar spherical harmonic, $\vb{e} = \sqrt{\varepsilon}\vb{E}/\mathcal{A}$ and $\vb{h} = \sqrt{\mu}\vb{H} / \mathcal{A}$ are the normalized  electric and magnetic fields. The integration in Eq.~\eqref{eq:AmjBmj} does not depend on the specific value of $r$, thus it can be arbitrary.

We decompose incident and scattered fields \eqref{eq:inc_plus_sc} using \eqref{eq:decomsition_general} separately with the set of coefficients $\{A^{\text{inc}}_{mj}, B^{\text{inc}}_{mj} \}$ and $\{A^{\text{sc}}_{mj}, B^{\text{sc}}_{mj} \}$ for incident and scattered fields, respectively.
For incident field we take $z_j(kr) = j_j(kr)$ spherical Bessel function, and for scattered field we take the spherical Hankel function of the first kind $z_j(kr) = h^{(1)}_j(kr)$, which has the correct asymptotic of a spherical outgoing wave at the infinity.
Once this decomposition is done, the surface integral in Eq.~\eqref{eq:torque_av} can be calculated analytically, and the $z$-component of the resulting torque $\vb{T} = (T_x, T_y, T_z)$ written in terms of contribution of spherical multipoles as
\begin{equation}
    T_z = \sum \limits_{mj} T_{mj,z} =
    \sum \limits_{mj}  m \hbar \left(\gamma^{\text{ext}}_{mj} - \gamma^{\text{sc}}_{mj}\right),
    \label{eq:torque_mj}
\end{equation}
with partial rates $\gamma^{\text{sc},\text{ext}}_{mj}~[1/\text{s}]$ are given by
\begin{align}
\begin{split}
    \hbar \gamma_{mj}^{\text{ext}} &= -T_0 j (j+1) \Re \left( A^{\text{inc}}_{mj} A^{\text{sc}*}_{mj} + B^{\text{inc}}_{mj} B^{\text{sc}*}_{mj}\right), \\
    \hbar \gamma_{mj}^{\text{sc}} &= T_0 j (j+1) \left( \abs{A^{\text{sc}}_{mj}}^2  + \abs{B^{\text{sc}}_{mj}}^2\right),
\end{split}
\label{eq:gamma_mj}
\end{align}
where $T_0  = |\mathcal{A}|^2 / (2 k^3)$ is the normalization torque with $\mathcal{A}$ being the dimensional constant in the decomposition Eq.~\eqref{eq:decomsition_general} of the incident field (see also Appendix~\ref{app:torque_norm}).
Explicit expression for the $T_x$ and $T_y$ components of the torque as well as the optical force can be found in Refs.~\cite{Barton1988,Barton1989}.
All the complexity of the problem is now hidden in the coefficients $A_{mj}$ and $B_{mj}$.
The dipole, quadrupole, and octupole contributions to the torque can be calculated as
\begin{align}
\label{eq:partial_torques}
\begin{split}
    \vb{T}^{\text{D}} &= \sum_{m=-1}^{1} \vb{T}_{m1} \equiv \vb{T}^{\text{e.D.}} + \vb{T}^{\text{m.D.}}, \\
    \vb{T}^{\text{Q}} &= \sum_{m=-2}^{2} \vb{T}_{m2} \equiv \vb{T}^{\text{e.Q.}} + \vb{T}^{\text{m.Q.}},  \\
    \vb{T}^{\text{O}} &= \sum_{m=-3}^{3} \vb{T}_{m3} \equiv \vb{T}^{\text{e.O.}} + \vb{T}^{\text{m.O.}}.
\end{split}
\end{align}
The higher multipolar contributions can found in a similar manner. Decomposition into the  electric or magnetic contributions (e.g. $\vb{T}^{\text{e.D.}}$ and $\vb{T}^{\text{m.D.}}$) is done by   keeping only $A^{\text{inc,sc}}_{mj}$ or $B^{\text{inc,sc}}_{mj}$ coefficients, respectively.

To provide a meaningful reference for comparison, we also introduce the concept of the waveplate torque limit~\cite{Friese1998Nat}
\begin{equation}
T^{\text{plate}} = \pi R^2 |\mathcal{A}|^2 / k,
\label{eq:Tplate}
\end{equation}
which is a maximum possible optical torque acting on a half-waveplate of area $\pi R^2$ that converts an incident RCP field into an outgoing LCP field with 100\% efficiency.
We note that we follow the standard normalization procedure when comparing area-dependent quantities, as optical forces and torques arise from the scattering of linear and angular momentum fluxes. This is fully analogous to the normalization of the scattering cross section by the geometrical cross section, \( \pi R^2 \). In this context, the optical force is typically normalized as \( F^{\text{geom}} = \pi R^2  |\mathcal{A}|^2 \), representing the pressure force exerted on an area equivalent to the geometrical cross section~\cite{Bekshaev2013OE,Almaas1995JOSAB}. Eq.~\eqref{eq:Tplate} is a direct analog for the torque in this context. See more on normalization in Appendix~\ref{app:torque_norm}.

Importantly, one can see from  Eq.~\eqref{eq:torque_mj} that the modes with zero angular momentum projection have no contribution to the spinning torque.
This can be attributed to the rotational symmetry of the $m=0$ eigenmode around the $z$-axis, making the particle to exhibit azimuthal symmetry in its optical response at resonance.
On the other hand, modes with higher total angular momentum projection give proportionately greater contribution to the optical spinning torque.

We also note that in principle, angular momentum in the incident field is not strictly necessary.
For instance, a linearly polarized plane wave can still produce a constant optical torque. 
The fundamental condition for a nonzero, wave-induced torque is that the combined \qqoute{incident field + particle} must be mirror-asymmetric, i.e., it is effectively \qqoute{chiral} in this context~\cite{Liu2010NN,Achouri2023ACS}. However, we will leave the \qqoute{pseudo-chirality} induced optical torque outside the scope of the current work.

The result Eq.~\eqref{eq:torque_mj} is general and was obtained without any assumptions on incident or scattered fields. 
We now exploit the multipolar expansion. We start with multipolar content of the \textit{incident field}, which is a standing electromagnetic wave formed by a two counter-propagating beams with opposite helicities:
\begin{equation}
    \sqrt{\varepsilon}\vb{E}_{\pm} = \mathcal{A} \begin{pmatrix}
        \cos k z \\ 
        \pm \iu \cos k z \\ 
        0
    \end{pmatrix}, \quad 
    \sqrt{\mu}\vb{H}_{\pm} = \mathcal{A} \begin{pmatrix}
        \pm \sin k z \\ 
        \iu \sin k z \\ 
        0
    \end{pmatrix},
    \label{eq:standing}
\end{equation}
where $\mathcal{A}$ is the magnitude, index \qqoute{$+$} corresponds to the spin angular momentum (SAM) density  $S_z > 0$, while index \qqoute{$-$} corresponds to $S_z < 0$. 
Standing wave Eq.~\eqref{eq:standing} has electric intensity anti-node at the origin $z=0$.
The normalized SAM density can be expressed as follows~\cite{Bliokh2017Aug}
\begin{equation}
    \bar{S}^{(\text{e})}_z = \cos^2 k z, \quad 
    \bar{S}^{(\text{m})}_z = \sin^2 k z, 
    \label{eq:SAM_standing_wave}
\end{equation}
where normalized values defined as
$\bar{S}^{(\text{e,m})}_z = \omega S^{(\text{e,m})}_z/U$,  with $U  = 1/4 \left( \varepsilon \abs{\vb{E}}^2  + \mu \abs{\vb{H}}^2\right) = |\mathcal{A}|^2/2$ being the energy density, and
$\vb{S} = (4\omega)^{-1} \Im \left( \varepsilon \vb{E}^* \times \vb{E} + \mu \vb{H}^* \times \vb{H} \right) \equiv \vb{S}^{(\text{e})} + \vb{S}^{(\text{m})}$ being the SAM density.
Standing wave \eqref{eq:standing} has no local phase gradients, i.e. zero local canonical momentum density, hence optical pressure force is zero~\cite{Toftul2024arXiv}. However, the gradients of electric $\varepsilon \abs{\vb{E}}^2/4$ and magnetic $\mu \abs{\vb{H}}^2 / 4$ parts of energy density are not zero and might provide a stable trapping in the nodes or anti-nodes.

Multipolar expansion of a standing wave~\eqref{eq:standing} with the help of Eq.~\eqref{eq:AmjBmj} is found to be 
\begin{align}
\begin{split}
    A^{\text{inc}(\pm)}_{mj} &= \delta_{1, j\bmod 2} \, \delta_{\pm 1, m} \,   \beta_j, \\
    B^{\text{inc} (\pm)}_{mj} &= \delta_{0, j\bmod 2} \, \delta_{\pm 1, m}  \, \beta_j,  
\end{split}
    \label{eq:standing_AB}
\end{align}
where $\beta_j =  \iu^{j+1}  \sqrt{4\pi \frac{2j+1}{j(j+1)}}$, $\delta_{ij}$ is the Kronecker delta, and factor $\delta_{0, j\bmod 2}$ leaves only even $j$, while $\delta_{1, j\bmod 2}$ leaves only odd $j$. For the standing wave which has a node of electric intensity at $z=0$ rather than anti-node, one has to change the $j$-parity in Eq.~\eqref{eq:standing_AB}.

For the case of an isotropic sphere, the scattering coefficients are
\begin{align}
\begin{split}
    A^{\text{sc}(\pm)}_{mj} &= - \delta_{1, j\bmod 2} \delta_{\pm 1, m} \beta_j a_j, \\
    B^{\text{sc}(\pm)}_{mj} &= - \delta_{0, j\bmod 2} \delta_{\pm 1, m}  \beta_j b_j,  
\end{split}
    \label{eq:sc_sphere_AB}
\end{align}
where $a_j$ and $b_j$ are the Mie scattering 
coefficients~\cite{bohren2008absorption} (see Appendix~\ref{app:mie}).  
Equation~\eqref{eq:sc_sphere_AB} shows that isotropic particles do not mix scattering channels in the vector spherical harmonics basis. If the incident field contains only a single multipole \( A^{\text{inc} (\pm)}_{mj} \) (\( B^{\text{inc} (\pm)}_{mj} \)), the scattered field will consist of \( A^{\text{sc} (\pm)}_{mj} \) (\( B^{\text{sc} (\pm)}_{mj} \)) with the same \( m \) and \( j \). In contrast, this behavior does not hold for particles with lower symmetry~\cite{Gladyshev2020PRB,Tsimokha2022PRB,Poleva2023PRB}.

Without losing any generality, further on we will consider only single polarization $\vb{E}_{+} \equiv \vb{E}_{\text{inc}}$.

\subsection{Lossy Mie-resonant cylinder}

In this section, we apply the multipolar expansion in order to observe  angular momentum transfer from field to cylindrical scatterer due to the \textit{enhanced absorption} at Mie resonances.
We start with configuration of a dielectric cylinder stably trapped in the counter-propagating beams.

\begin{figure}
    \centering
    \includegraphics[width=\linewidth]{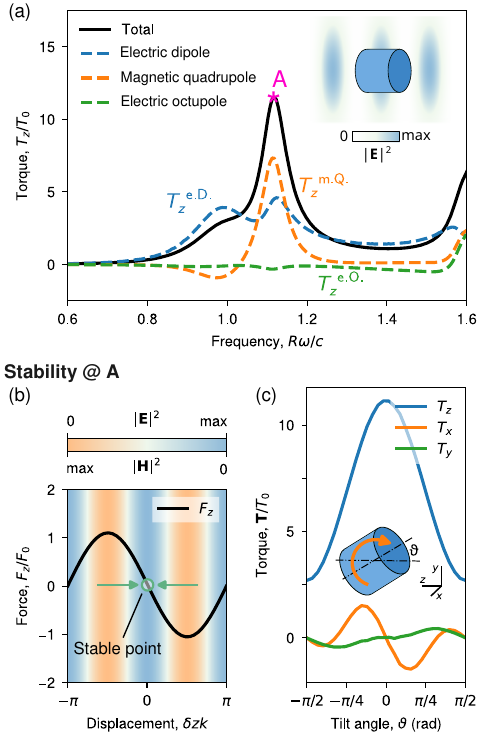}
    \caption{ \textbf{Characteristics of spinning optical torque.} (a) Spectrum of the total torque \eqref{eq:torque_av} and  multipolar expansion \eqref{eq:partial_torques} for a cylinder positioned at the standing wave's anti-node. (b,c) Positional and rotational stability analyses of the cylinder at the point of the maximum torque in the considered spectrum range. Simulation parameters: cylinder permittivity $\varepsilon = 16 + \iu 0.5$, radius $R = 100~\text{nm}$, height $H = 161~\text{nm}$. Results are normalized by $F_0 =  |\mathcal{A}|^2 / (2k^2)$ and $T_0  = |\mathcal{A}|^2 / (2 k^3)$.}
    \label{fig2:resonant_torque_2beams}
\end{figure}

The counter-propagating geometry of illumination compensates the  radiation pressure such that trapping is solely determined by the gradient force~\cite{Toftul2024arXiv,Lepeshov2023Jun}. 
We fix the  cylinder geometry, radius  $R = 100~\text{nm}$, and the  height $H = 161~\text{nm}$ (Fig.~\ref{fig1:2beams_am}), giving the aspect ratio of $H/R = 1.61$, the permittivity of material is set to be equal to $\varepsilon=16+\iu 0.5$. 
The amount of non-radiative material losses, i.e. the imaginary part of the permittivity $\Im(\varepsilon)$, is chosen to satisfy the conditions for the \textit{critical coupling regime} for a sphere in this spectral range. Critical coupling occurs when intrinsic material losses and radiative losses are balanced~\cite{Cheng2021ACSPhot,Rahimzadegan2017PRB}, 
leading to maximal energy absorption~\cite{Ruan2010PRL,Suh2004IEEE}. 
Indeed, since high Ohmic losses will suppress the Q-factor and low Ohmic losses do not provide enough absorption, there exists an optimal value ensuring the maximal absorption \cite{Zograf2021AOP}. 
Close to the light  frequency of $\omega R/c\approx 1$ the cylinder has pronounced Mie resonances. 
Surprisingly, for this setup we find two distinct stable trapping configurations: (i) trapping in the anti-node (maximum of the electric field intensity), and (ii) trapping in the node (minimum of the electric field intensity). 
Figure~\ref{fig2:resonant_torque_2beams} demonstrates the anti-node trapping condition while the node trapping is discussed in Appendix~\ref{app:node_trapping}. 

One can see in Fig.~\ref{fig2:resonant_torque_2beams}(a) that the optical torque is  enhanced at the resonances which have complex multipolar origin both in anti-node and node trapping conditions. The torque is normalized over $T_0$.  
The multipolar content is defined by the particular multipolar structure of the resonant mode. Moreover, some of the multipolar torque components can be negative, 
even though the total torque remains positive.
This originates from the fact that different multipolar channels in the scattering field are not orthogonal and can be mixed by  cylindrical symmetry of the object~\cite{Gladyshev2020PRB,Poleva2023PRB}.  
This result was obtained by direct full wave numerical simulations of light scattering on the cylinder followed by integration of the stress tensor according to Eq.~\eqref{eq:torque_av}. The multipolar contributions to the torque were obtained by first making the multipolar decomposition of the scattered field using Eq.~\eqref{eq:AmjBmj} and then by using the expression Eq.~\eqref{eq:partial_torques} separate multipolar contributions are obtained.

In order to observe stable optical trapping it is  sufficient to have \textit{positional stability}, i.e. the center of mass of the particle must stand in a stable equilibrium position. 
We calculate the total spinning torque using Eq.~\eqref{eq:torque_av} and its multipolar contributions using Eqs.~\eqref{eq:partial_torques}, based on a numerical scattering problem solved in COMSOL Multiphysics (see Appendix~\ref{app:numerical_methods}).
One can see that the anti-node stability conditions can be achieved at the local torque peak value [point A in Fig.~\ref{fig2:resonant_torque_2beams}(a)].
There is a restoring force keeping the particle in the field maximum/minimum as shown in Fig.~\ref{fig2:resonant_torque_2beams}(b). However, for the case of non-spherical particle rotation it is also very important to have stability regarding the particle tilt, otherwise any deviation of the cylinder axis from the beam axis can lead to complex dynamics of the particle~\cite{Brzobohaty2015SR, Brzobohaty2015OE}. 
Surprisingly, the  stable point also demonstrates the \textit{tilt stability}. Fig.~\ref{fig2:resonant_torque_2beams}(c) shows that rotation of the particle around the axis orthogonal to the beam axis ($x$-axis for certainty) results in the appearance of the torque rotating the particle in the opposite direction. Eventually, the circularly polarized  standing wave  provides the conditions for observing stable rotation of a cylindrical particle.

\subsection{Anisotropic Mie-resonant cylinder}

\begin{figure*}
    \centering
    \includegraphics[width=\linewidth]{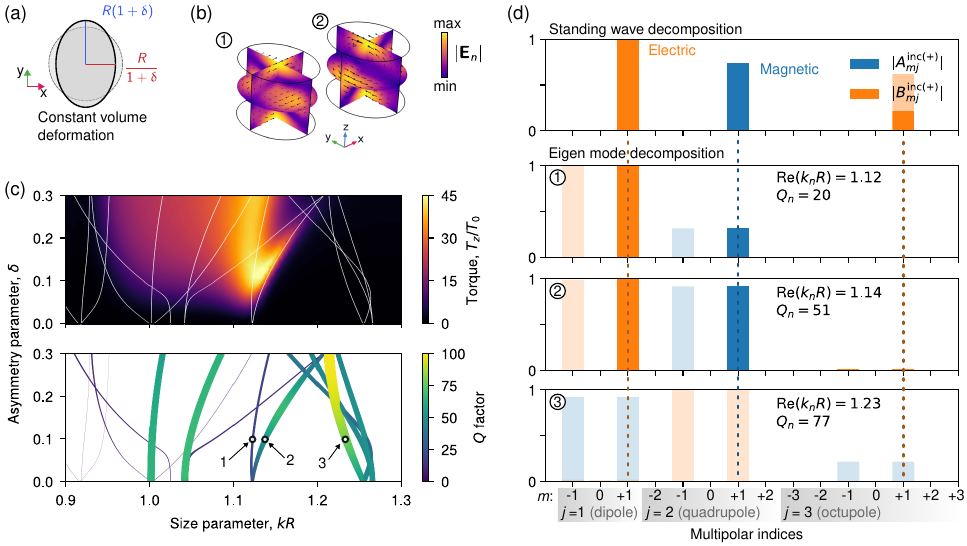}
    \caption{\textbf{Effect of deformation.} 
        (a) Illustration of the constant volume deformation such that $V = \pi R^2 H = \operatorname{const}(\delta)$.
        (b) Distributions of the electric field of two well excited eigenmodes.
        (c) 2D parametric maps of size parameter (dimensionless frequency) vs asymmetry parameter. Top: normalized values of the optical torque. Bottom: eigenmodes of the cylinder. The permittivity of the cylinder is chosen to be $\varepsilon = 16$.
        (d) Multipolar expansion of the standing wave based on Eq.~\eqref{eq:standing_AB} and numerically calculated coefficients of the three different eigenmodes using Eq.~\eqref{eq:AmjBmj}. All the values are normalized by the $\operatorname{max}(|A_{mj}|, |B_{mj}|)$.
    }
    \label{fig3:torque_eigen_shift}
\end{figure*}

The second mechanism responsible for the optical rotation of the object is related to the particle anisotropy, which breaks the cylindrical symmetry of the object and, thus, results in the mixing the scattering channels with different angular momentum. This effect has already been proposed for observing effective rotation of particles of various shapes~\cite{Friese1998Nat,Xu2017PRA,Shao2015ACSNano,Higurashi1997JAP}, including  negative rotation~\cite{Hakobyan2014NP, Chen2014SR, Han2018NC, Diniz2019OE, Shi2022Jul, Qi2022NL}. 
The torque components $T_{mj,z}$ in Eq.~\eqref{eq:torque_mj} consist of extinction and scattering contributions. According to Eq.~\eqref{eq:gamma_mj}, the extinction rate $\gamma^{\text{ext}}_{mj}$ shares the same set of azimuthal numbers $m$ as the incident field, referred to as $m_{\text{inc}}$. For structures with $m_s$-fold rotational symmetry (where the $z$-axis is the axis of symmetry), the set of $m$ in the scattering contribution ($m_{\text{sc}}$) is defined by the selection rule $m_{\text{sc}} = m_{\text{inc}} + n m_s$, with $n$ being an integer~\cite{Chen2014SR}. By combining the extinction and scattering terms with $n=0$, we derive an alternate form of Eq.~\eqref{eq:torque_mj}:
\begin{equation} 
    T_z = \sum \limits_{m_{\text{inc}}} m_{\text{inc}} \hbar \gamma^{\text{abs}}_{m_{\text{inc}}} - \sum \limits_{m=-\infty}^{\infty} \delta_{m m_{\text{sc}}} m \hbar\gamma^{\text{sc}}_{m}, 
    \label{eq:T_m_gamma} 
\end{equation}
where $n = \pm1, \pm2, \dots$; Kronecker symbol $\delta_{m m_{\text{sc}}}$ equals $1$ if there exists at least one combination of $m_{\text{inc}}$ and $n$ that provides such $m$, and $0$ otherwise. Notably, $n=0$ values are excluded in the second term to express absorption-related torque explicitly in the first term. The scattering and absorption rates in \eqref{eq:T_m_gamma} are connected with the VSH angular momentum basis as $\gamma^{\text{sc},\text{abs}}_m = \sum \limits_{j=|m|}^{\infty}\gamma_{mj}^{\text{sc},\text{abs}}$ and $\gamma^{\text{abs}}_{mj} = \gamma^{\text{ext}}_{mj} - \gamma^{\text{sc}}_{mj}$.
For the case of small dipole scatterers one can explicitly split the optical torque into the absorption and anisotropic parts~\cite{Toftul2024arXiv,Toftul2023PRL}. In this situation the anisotropic part appears quite naturally as the difference between the polarizability in two main directions.

In order to trace the behavior of Mie resonators, we considered a cylinder with introduced asymmetry parameter $\delta$ which deforms it into an elliptical cylinder with $m_s = 2$ as shown in Fig.~\ref{fig3:torque_eigen_shift}(a). The deformation preserves the volume of the particle. 
We also considered purely lossless system with $\varepsilon=16$ where the Ohmic absorption is fully suppressed for the moment. The spectral dependence of the optical torque \eqref{eq:torque_av} on the asymmetry parameter $\delta$ is shown in Fig.~\ref{fig3:torque_eigen_shift}(c, top panel).  One can see that the strong resonant increase of the optical torque is observed  under illumination of the elliptical cylinder by a circularly polarized beam. This resonance corresponds to the excitation of high-order  Mie modes and results in drastic enhancement of the optical torque reaching the value of $45 T_0 \simeq 5.7 T^{\text{plate}}$, where $T^{\text{plate}}$ is given by Eq.~\eqref{eq:Tplate}. Changing the shape of the cylinder base from circular to elliptical also defines the spectral shift of the resonances  and the $Q$-factors as it is shown in Fig.~\ref{fig3:torque_eigen_shift}(c, bottom panel). These plots correspond to the anti-node trapping geometry shown in Fig.~\ref{fig2:resonant_torque_2beams}(a). Eigen modes field distributions for $A_{mj}$ and $B_{mj}$ calculations using Eq.~\eqref{eq:AmjBmj}, frequencies, and  $Q$-factors are calculated using COMSOL Multiphysics eigenfrequency solver (see Appendix~\ref{app:numerical_methods}).

The resonant increase of the optical torque is connected mostly to the excitation of modes 1 and 2, their field distribution is also shown in Fig.~\ref{fig3:torque_eigen_shift}(b). At the same time, the other modes with even higher $Q$-factor do not contribute to the optical torque, such as  mode 3, for instance. It turns out, that these modes are uncoupled with the incident field. Most clearly it can be illustrated with their multipole content. Indeed, the multipole content of the circularly polarized standing wave Eq.~\eqref{eq:standing_AB}  is shown in the top panel of Fig.~\ref{fig3:torque_eigen_shift}(d). The multipolar  content of modes 1--3 having dipole-quadrupole character is also shown in the panels. One can see that the modes 1 and 2 \textit{match} with the multipoles in the incident field having dipole and quadrupole components with $m=+1$. At the same time, mode 3 is \textit{uncoupled} to the incident field due to electric-magnetic modes mismatch.

At this point, it is illustrative to compare the optical torque on an elliptical cylinder with the optical torque acting over a spherical particle of the same radius, where all the multipolar channels stay  orthogonal and independent.  From the general expression \eqref{eq:torque_mj} and coefficients \eqref{eq:standing_AB}--\eqref{eq:sc_sphere_AB}, we find that the torque acting on an isotropic sphere in a standing wave \eqref{eq:standing} is
\begin{equation}
    T^{(\pm)}_z = \pm 2 T_0 k^2 \left( \sum \limits_{\text{odd}~j} \sigma^{\text{abs,e}}_{j} + \sum \limits_{\text{even}~j} \sigma^{\text{abs,m}}_{j} \right),
    \label{eq:torque_sphere_standing}
\end{equation}
where $\sigma^{\text{abs,e}}_{j} = 2\pi k^{-2} (2j+1) \left(\Re(a_j) - \abs{a_j}^2\right)$ is the partial electric absorption cross section and $\sigma^{\text{abs,m}}_{j} = 2\pi k^{-2} (2j+1) \left(\Re(b_j) - \abs{b_j}^2\right)$ is the magnetic absorption cross section.
For a sphere in an electric field node one should swap $\sum_{\text{even}~j} \leftrightarrow \sum_{\text{odd}~j} $.  For a sphere we can find a limit on the partial torque contributions $T_{mj,z}$ from Eq.~\eqref{eq:torque_mj}.
At the resonance the potential limits for Mie coefficients are $\Re(a_j) - |a_j|^2 = 1/4$ and $\Re(b_j) - |b_j|^2 = 1/4$~\cite{Miroshnichenko2018PRL}. 
As a result,  the maximum possible partial absorption cross section is $\operatorname{max} \left(\sigma^{\text{abs,e}}_{j} \right) = \operatorname{max} \left(\sigma^{\text{abs,m}}_{j} \right) = \pi (2j+1) / (2k^2)$. Substitution of this result to  Eq.~\eqref{eq:torque_sphere_standing} leads to standing wave \textit{partial torque limit} for an isotropic spherical particle caused by absorption mechanism:
\begin{equation}
    \operatorname{max} \left(|T_{1j,z}| \right) =    \pi (2j+1) T_0.
    \label{eq:limit}
\end{equation}
Eq.~\eqref{eq:limit} defines the $j$-pole optical torque limit, i.e. $j=1$ provides dipole limit, $j=2$ provides quadrupole limit, etc. We note that in a propagating plane wave this limit is twice bigger since for the same $j$ there are both electric and magnetic contributions. 
This result correlates with known fundamental limits on optical torque~\cite{Rahimzadegan2017PRB}. Alternatively, it is related to critical coupling condition when the radiative and non-radiative losses of a particular mode are balanced \cite{Grigoriev2015ACSPhot, Miroshnichenko2018PRL,Ryabov2022Nanophotonics,Rahimzadegan2017PRB, Cheng2021ACSPhot,Suh2004IEEE,Ruan2010PRL}.
{Equation \eqref{eq:limit} can be interpreted as follows: 
the maximum absorption torque occurs when the angular momentum of a multipole in the incident wave, with total angular momentum $\hbar j$, is fully absorbed by the particle.
However, the partial torque on non-spherical particles can potentially exceed this limit by a mechanism of \textit{channel mixing}.
Particles with lower symmetries mix various multipolar channels, in principle infinitely many~\cite{Krasikov2021PhysRevAppl,Tsimokha2022PRB,Poleva2023PRB,Gladyshev2020PRB}, and such simple limiting condition for each channel is not valid anymore, however the conservation of the total angular momentum is preserved.
}

\begin{figure}
    \centering
    \includegraphics[width=\linewidth]{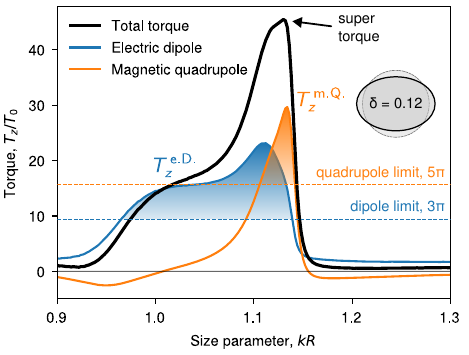}
    \caption{\textbf{Supertorque.} Optical torque on an anisotropic particle with $\delta = 0.12$ located in the anti-node of a standing wave. The occurrence of the optical supertorque is explained by the spectral overlap of dipole and quadrupole torque components, each of which exceeds the single channel limit due to the mode mixing. The maximum achieved value is $T_z \simeq 45 T_0 \simeq 5.7 T^{\text{plate}}$. Total torque is calculated by Eq.~\eqref{eq:torque_av} and its multipolar decompositions are calculated using Eqs.~\eqref{eq:partial_torques}.}
    \label{fig:beyond_limit_anisotropic}
\end{figure}

One approach to achieve channel mixing involves utilizing \textit{super-absorption} regimes, where multiple independent absorption channels are engineered to exhibit resonances at the same wavelength~\cite{Ladutenko2015N}.
An alternative way to break the limit \eqref{eq:limit} is to increase the scattering contribution in Eq.~\eqref{eq:T_m_gamma}, which can be achieved only by the rotational symmetry breaking.
This method proves to be significantly more effective: for a lossless anisotropic cylinder with an asymmetry parameter of $\delta = 0.12$, we achieved a dipole torque exceeding the dipole limit of $T_z = 3\pi T_0$ by more than two times (see Fig.~\ref{fig:beyond_limit_anisotropic}). The maximal torque for the spherical particle of the same radius is shown with horizontal dashed line.  Moreover,  in the same spectral range, the quadrupole torque also exceeds the corresponding limit for spherical particle. This increase of the total optical torque over the predicted limits we refer to as ``{\textit{supertorque}}'' in similar manner to super-absorption and super-scattering  regimes. Here the partial multipolar torques are obtained within Eq.~\eqref{eq:partial_torques}. We also need to note that as  clear sign of different multipole channels mixture, there is  a region of negative partial quadrupole torque at $kR\approx0.95$ seen in Fig.~\ref{fig:beyond_limit_anisotropic}. It corresponds to energy exchange from the dipole to  quadrupole scattering channels. The total torque as a sum of partial torques stays strictly positive at the same time.

\subsection{Asymmetry and losses: Dual contribution}

\begin{figure}
    \centering
    \includegraphics[width=\linewidth]{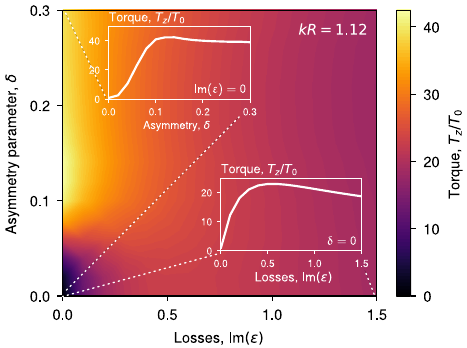}
    \caption{
        \textbf{Asymmetry and losses.} Search for the optimal ratio between internal losses and asymmetry of the rotated cylindrical nanoparticle. Both the asymmetry and losses at the same time bring the possibility of spinning torque but destroy the mode $Q$-factor.
    }
    \label{fig:losses_vs_delta}
\end{figure}

The two mechanisms, absorption and asymmetry induced, of angular wave-to-particle AM transfer may simultaneously add to the optical torque. To analyze that, we have plotted the optical torque dependence on the imaginary part of the permittivity and on the anisotropy parameter $\delta$, while preserving constant $kR = 1.12$ (Fig.~\ref{fig:losses_vs_delta}). One can see that for nearly round cylinders $\delta\approx0$ the optical torque is governed by  losses and has maximum at moderate level which is a sign of maximized absorption due to the critical coupling condition point. The increase of the anisotropy in the absence of losses and for small values of $\delta$ provide increasing optical torque $T_z\sim \delta^2$  according to perturbation theory (see Supplementary Materials of Ref.~\cite{Toftul2023PRL}). So one can conclude, that small amount of losses and anisotropy both add to optical torque. However, this is not generally true for larger values. Indeed, for  large losses increase of the particle anisotropy can even reduce the optical torque and, thus, two mechanisms can not be generally separated from each other. It is also interesting that the maximal torque weakly depends on the anisotropy for large $\delta$ and small $\Im(\varepsilon)$, which is probably due to the particular mode structure, as its $Q$ factor and multipole content are conserved all over the considered parameters range.

\subsection{Experimental estimations}

Here we estimate the potential rotation frequencies of Mie resonant nanoparticles in air. 
The viscosity torque damps the rotation of the particle due to the viscous interaction with the gas molecules, and is proportional (and opposite to) the rotation frequency of the particle: $\vb{T}^{\text{drag}} = - \nu \vb{\Omega}$, where $\nu$ is the proportionality constant. 
At low Reynolds numbers, for the case of a cylinder rotating in a liquid or gas with dynamical viscosity $\eta$ the proportionality constant  is $\nu  =  2\pi \eta R^2 H$~\cite{kim2013microhydrodynamics}.
The drag torque scales quadratically with asymmetry parameter $\delta$ introduced in Fig.~\ref{fig3:torque_eigen_shift}(a), hence we write
\begin{equation}
    \vb{T}^{\text{drag}} = - 2\pi \eta R^2 H  (1 + \beta \delta^2) \vb{\Omega} ,
    \label{eq:drag}
\end{equation}
where $\beta$ is an empirical dimensionless geometrical factor, which can be analytically calculated for the case of an ellipsoid~\cite{kim2013microhydrodynamics}.
For a typical experimental setup with $\lambda = 1550$~nm and laser power of around $100$~mW, we calculate that for the objective lens with $\mathrm{NA}=0.8$ the normalization value of the torque is $T_0 \approx 2 \times 10^{-17}$~\unit{N}$\cdot$\unit{m}. 
Assuming air as the host medium, at normal conditions its dynamic viscosity is $\eta \approx 2 \times 10^{-5}$~\unit{Pa}$\cdot$\unit{s}~\cite{Gururaja1967J.Chem.Eng.Data}. 
In equilibrium, the drag torque Eq.~\eqref{eq:drag} is fully compensated by the optical torque Eq.~\eqref{eq:torque_av}, $\vb{T}^{\text{drag}} + \vb{T} = 0$. This leads to the approximation of the \textit{non-resonant} rotation frequency for $\delta = 0$:
\begin{equation}
    \frac{\Omega_{0}}{2\pi} = T_0/(2\pi \nu) \approx 2\times 10^{7}~\text{Hz}.
\end{equation} 
For the resonant lossy cylinder made of high index material, i.e. with relative permittivity of $\varepsilon \approx 16$ (e.g. Si in vacuum), the rotation frequency can increase by up to 13 times [Fig.~\ref{fig2:resonant_torque_2beams}(a)], while resonant lossless particle provides a 45 times enhancement (Fig.~\ref{fig:beyond_limit_anisotropic}) and could almost reach the GHz rotation frequency:
\begin{equation}
    \frac{\Omega^{\text{lossy}}}{2\pi} \approx 3 \times 10^{8}~\text{Hz}, \qquad 
    \frac{\Omega^{\text{anis}}}{2\pi} \approx 9 \times 10^{8}~\text{Hz}.
    \label{eq:estimations}
\end{equation}
This analysis provides only general estimates. We acknowledge that $\beta$ values may vary greatly, and at high rotation frequencies  drag torque Eq.~\eqref{eq:drag} may lose validity.

Finally, we emphasize that the values in Eq.~\eqref{eq:estimations} were estimated under ambient conditions. In a regime where the mean free path of gas molecules is much bigger than the particle diameter, the spinning frequency is inversely proportional to the gas pressure, $\Omega/2\pi \propto 1/p_{\text{gas}}$ ~\cite{Reimann2018PRL,Zielinska2024PRL}.
Consequently, even higher enhancements are expected in high vacuum. However, such extreme spinning frequencies of nanoparticles are not yet well studied and the exact mechanism which limits the maximum spinning frequency remains unknown. 
Thus, Mie resonant \qqoute{supertorque} could potentially provide a valuable platform for studying the limits of high frequency spinning.
Finally, this resonant enhancement can be further combined with material anisotropy, as seen in vaterite or calcite nanoparticles~\cite{Noskov2018Nanoscale,Barhum2024ChemEngJ}, which requires further investigation.

\section{Conclusions}

We have studied the general mechanism of the generation of the optical torque due to light scattering on Mie-resonant cylindrical subwavelength particles.  We have explored the regimes of stable trapping and stable rotation of a dielectric subwavelength cylinder in the anti-nodes of either electric or magnetic fields. We have predicted and demonstrated theoretically that in the supertorque regime one can expect a ten-fold enhancement of an optical torque compared to the non-resonant case,  with its magnitude overcoming the corresponding limit for an isotropic sphere. 
The \textit{supertorque} effect differs from the well-known super-absorption regime, as it contains a broader range of phenomena and specifically emphasizes the angular momentum imbalance in scattering, which can arise from both \textit{absorption} and \textit{asymmetry}.

In addition, we have observed that, in a perturbative regime, both anisotropy and optical losses independently contribute to the enhancement of the resonant optical torque acting on a Mie cylinder. At the same time, for high-loss materials and strongly anisotropic structures, the contribution of those two effects into the optical torque can not be separated, and the overall effect has a complex character. We believe our results will be important for increasing the level of optical control and manipulation over highly resonant nanoscale objects,  and they can be potentially applied in biophysics and nanochemistry, as well as for the study of the fundamental interactions within the field of levitodynamics.

\section*{Acknowledgement}
This work was supported by the Australian Research Council (Grant No. DP210101292) and the International Technology Center
Indo-Pacific (ITC IPAC) via Army Research Office (contract
FA520923C0023).  A part of the theoretical analysis was supported by the Russian Science Foundation (Grant 20-72-10141). The work was also supported by the Federal Academic Leadership Program Priority 2030.

\bibliography{refs}

\onecolumngrid
\newpage
\appendix 

\section{Numerical methods}
\label{app:numerical_methods}

Numerical simulations were performed in the Wave Optics module of COMSOL Multiphysics. The near-field distributions, resonant wavelengths, and $Q$ factors are simulated using the eigenmode solver. The scattering problem was simulated using frequency domain solver. The background field was introduced to the model according to the Eq.~\eqref{eq:standing}.

The data that support the findings of this study, including Python notebooks and COMSOL data files used to generate the plots in the paper, are openly available in Zenodo repository~\cite{Toftul2024Zenodo}.

\section{Vector spherical harmonics}
\label{app:VSH}

Electric $\vb{N}_{mj}$ and magnetic $\vb{M}_{mj}$ vector spherical harmonics used throughout this paper in spherical coordinates $(r, \vartheta, \varphi)$ are given by
\begin{align}
    \vb{M}_{mj} =& \ort{\vartheta}  \frac{\iu m}{\sin \vartheta} z_j Y_{mj} - \ort{\varphi} z_j  \left(Y_{mj}\right)^{\prime}_{\vartheta} ,\\
    \vb{N}_{mj} =& \ort{r} j(j+1) \frac{z_j}{kr} Y_{mj} 
    + \ort{\vartheta} \frac{1}{kr}  \left( r z_j \right)^{\prime}_{r} \left(Y_{mj}\right)^{\prime}_{\vartheta} + \ort{\varphi} \frac{\iu m}{kr \sin \vartheta} \left( r z_j \right)^{\prime}_{r} Y_{mj},
\end{align}
where $z_j = z_j(kr)$ is the radial spherical function, $Y_{mj} = Y_{mj}(\vartheta, \varphi)$ is the scalar spherical harmonic, and $\ort{r}$, $\ort{\vartheta}$, $\ort{\varphi}$ are the unit orts of spherical coordinates. $\vb{N}_{mj}$ and $\vb{M}_{mj}$ are solutions of the Maxwell equation and satisfy
\begin{equation}
    \grad \times \vb{N}_{mj} = k \vb{M}_{mj}, \quad 
    \grad \times \vb{M}_{mj} = k \vb{N}_{mj}.
\end{equation}
See more in SM of Ref.~\cite{Toftul2023PRL}.

\section{Mie scattering coefficients}
\label{app:mie}

The problem of plane wave scattering by an isotropic sphere can be solved exactly. This solution is commonly referred to as the Mie solution in the name of Gustav Mie~\cite{Mie1908AP}. The electromagnetic Mie coefficients for a sphere of radius $a$ are given by~\cite{bohren2008absorption}:
\begin{equation}
    a_{n} = \frac{\sqrt{\bar{\varepsilon}}\, \psi_n(k_{p} a) \psi^{\prime}_n(k a) - \sqrt{\bar{\mu}}\, \psi_n(k a) \psi^{\prime}_n(k_{p} a)}{\sqrt{\bar{\varepsilon}}\, \psi_n(k_{p} a) \xi^{\prime}_n(k a) - \sqrt{\bar{\mu}}\, \xi_n(k a) \psi^{\prime}_n(k_{p} a)}, 
    \qquad 
    b_{n}  = \frac{\sqrt{\bar{\mu}}\, \psi_n(k_{p} a) \psi^{\prime}_n(k a) - \sqrt{\bar{\varepsilon}}\, \psi_n(k a) \psi^{\prime}_n(k_{p} a)}{\sqrt{\bar{\mu}}\, \psi_n(k_{p} a) \xi^{\prime}_n(k a) - \sqrt{\bar{\varepsilon}}\, \xi_n(k a) \psi^{\prime}_n(k_{p} a)}, 
    \label{eq:Mie_coef}
\end{equation}
where $\psi_n(x) = x j_n(x)$ and $\xi_n(x) = x h_n^{(1)}(x)$ are the Riccati-Bessel functions, $j_n$ is the spherical Bessel function, $h_n^{(1)}$ is the spherical Hankel function of the first kind, the prime denotes derivative with respect to the argument, $\bar{\varepsilon} = \varepsilon_p /\varepsilon$ and $\bar{\mu} = \mu_p /\mu$ are the relative permittivity and permeability, and  $k_p = \omega \sqrt{\varepsilon_p \mu_p}$ is the wavenumber inside the sphere.

\section{Node stability analysis}
\label{app:node_trapping}

In the main text it was shown that cylinder can have positional and tilt stability in the anti-node of the standing wave. Here we show that it is also possible for the node trapping scheme as well.

\begin{figure}
    \centering
    \includegraphics[width=0.5\linewidth]{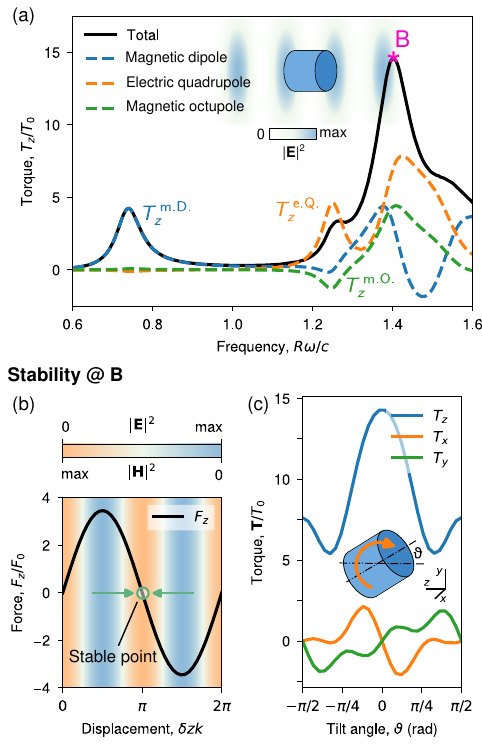}
    \caption{(a) Spectrum of the spinning optical torque and its multipolar expansion on a cylinder which is located in the electric node of the standing wave. Positional (b) and rotational (c) stability analyses of the cylinder at the point of the maximum torque in the considered spectrum range. Simulation parameters: cylinder permittivity $\varepsilon = 16 + \iu 0.5$, radius $R = 100~\text{nm}$, height $H = 161~\text{nm}$. Results are normalized by $F_0 =  |\mathcal{A}|^2 / (2k^2)$ and $T_0  = |\mathcal{A}|^2 / (2 k^3)$.}
    \label{fig:node_stability}
\end{figure}

Cylinder placed in the electric node is going to have different multipolar content in the scattered field [Fig.~\ref{fig:node_stability}(a)], somewhat akin analytical solution for a spherical particle. At the frequency $R \omega / c \approx 1.4$ (point B) torque reaches maximal value of around $T_z/T_0 \approx 15$. Panels (b) and (c) in Fig.~\ref{fig:node_stability} show positional and tilt stability at point B.

\section{Torque normalization}
\label{app:torque_norm}

One might be inclined to compare a cylinder with a sphere of equivalent volume.
However, we deliberately avoided such a comparison.
All plots present torque values normalized by the $T_0 = |\mathcal{A}|^2 / (2k^3)$, which is just a dimensional constant in the equations and depends only on the incident wave amplitude and its wave number $k$. For example, once the electric dipole contribution of the torque exceeds
\begin{equation}
    \frac{T_z^{\text{e.D.}}}{T_0} > 3 \pi
\end{equation}
one has achieved a ``supertorque'' regime by \textit{definition}. The concept of equivalent volume is not required in this reasoning. 

We note that we follow the standard normalization procedure when comparing area-dependent quantities, as optical forces and torques arise from the scattering of linear and angular momentum fluxes. This is fully analogous to the normalization of the scattering cross section by the geometrical cross section, \( \pi R^2 \). In this context, the optical force is typically normalized as \( F^{\text{geom}} = \pi R^2  |\mathcal{A}|^2 \), representing the pressure force exerted on an area equivalent to the geometrical cross section~\cite{Bekshaev2013OE,Almaas1995JOSAB}. 
In this context, the geometrical limit for the torque, analogous to the normalization of optical force, is given by the waveplate limit introduced in the main text:
\begin{equation}
    T^{\text{plate}} = \pi R^2 |\mathcal{A}|^2 / k.
\end{equation}

\end{document}